\documentclass[sigconf,table]{acmart}

\usepackage{tikz}
\usetikzlibrary{bayesnet}
\usepackage{booktabs} 
\usepackage{tabularx} 
\usepackage[english]{babel}
\usepackage{mathrsfs}

\usepackage{multirow}
\usepackage{multicol}
\usepackage{array}
\usepackage{arydshln}

\usepackage{amsmath}
\usepackage{booktabs}
\usepackage[inline]{enumitem}
\usepackage{siunitx}
\usepackage{stackengine}
\usepackage{extarrows}

\usepackage{caption}
\usepackage{subcaption}

\usepackage{soul}
\usepackage{flushend}

\usepackage{footnote}
\makesavenoteenv{tabular}
\makesavenoteenv{table}

\usepackage{xspace} 

\newcolumntype{L}[1]{>{\raggedright\arraybackslash}p{#1}}
\newcolumntype{C}[1]{>{\centering\arraybackslash}p{#1}}
\newcolumntype{R}[1]{>{\raggedleft\arraybackslash}p{#1}}

\copyrightyear{2021}
\acmYear{2021}
\setcopyright{acmlicensed}\acmConference[SIGIR '21]{Proceedings of the 44th International ACM SIGIR Conference on Research and Development in Information Retrieval}{July 11--15, 2021}{Virtual Event, Canada}
\acmBooktitle{Proceedings of the 44th International ACM SIGIR Conference on Research and Development in Information Retrieval (SIGIR '21), July 11--15, 2021, Virtual Event, Canada}
\acmPrice{15.00}
\acmDOI{10.1145/3404835.3462805}
\acmISBN{978-1-4503-8037-9/21/07}

\author{Rosie Jones$^{*1}$, Hamed Zamani$^{*2}$, Markus Schedl$^{3}$, Ching-Wei Chen$^{1}$, Sravana Reddy$^{1}$, Ann Clifton$^{1}$, Jussi Karlgren$^{1}$, Helia Hashemi$^{2}$, Aasish Pappu$^{1}$, Zahra Nazari$^{1}$, Longqi Yang$^{4}$, Oguz Semerci$^{1}$, Hugues Bouchard$^{1}$, Ben Carterette$^{1}$} \thanks{$^*$ Equal contribution.}

\affiliation{
  \institution{$^{1}$ Spotify \quad $^{2}$ University of Massachusetts Amherst \quad $^{3}$ Johannes Kepler University Linz \quad $^{4}$ Microsoft}
}
\affiliation{
  \institution{$^{1}$ \{rjones, cw, sravana, aclifton, jkarlgren, aasishp, zahran, oguz, hb, benjaminc\}@spotify.com \\ $^{2}$ \{zamani, hhashemi\}@cs.umass.edu \quad $^{3}$ markus.schedl@jku.at \quad $^{4}$ longqi.yang@microsoft.com}
}

\begin{document}
\title{Current Challenges and Future Directions in\\Podcast Information Access}

\begin{abstract}
Podcasts are spoken documents across a wide-range of genres and styles, with growing listenership across the world, and a rapidly lowering barrier to entry for both listeners and creators. The great strides in search and recommendation in research and industry have yet to see impact in the podcast space, where recommendations are still largely driven by word of mouth. In this perspective paper, we highlight the many differences between podcasts and other media, and discuss our perspective on challenges and future research directions in the domain of podcast information access.
\end{abstract}



\maketitle

\section{Introduction}  
\label{sec:intro}




With the click of a button, virtually any person with a smartphone and a podcast app such as Anchor\cite{anchor} or Podbean \cite{podbean} 
can record, edit, and publish a podcast to all the leading audio streaming platforms. As podcasting has greatly reduced the cost of producing and distributing audio content, there has been a massive increase in the number of podcasts: as of January 2021, the podcast search engine Listen Notes\cite{listennotes:stats} lists over 1.9M podcast shows, and over 90M episodes hosted on public RSS servers - more than double the number from Dec 2019. 

The listening audience for podcasts has kept pace, growing to a critical mass in recent years. Edison Research reports\cite{infinitedial2020} that podcast listening grew from 11\% of the US population in 2006,
%
to 55\% in 2020. The same report found that in the US, weekly listeners spent an average of 6 hours and 39 minutes listening to podcasts. Growth in the segment is expected to continue at a rapid pace.
According to PwC\cite{pwcoutlook},
global podcast listenership was around 600M in 2019, and is projected to grow to 1.5B by 2024. PwC also project that podcast advertising will approach \$3.5B in 2024, an increase from roughly \$1B in 2019. 

With
a massive amount of podcast content, and an eager audience,
there are many open questions around how best to provide access to this information. Past work in spoken document retrieval \cite{garofolo2000trec} is based on news corpora, while podcasts come in many disparate genres. Recommendations from friends and family \cite{Edison2019Consumer} remain in the top-three ways people find podcasts, while non-podcast-listeners in the same study say that they don't know how to find a podcast.
We believe that existing technology is insufficient for providing efficient access to podcasts, which necessitates further research on the topic.

In this paper, we lay out the challenges of podcast information access, and highlight areas which are important for further research.
We introduce the basic characteristics of podcasts and highlight their similarities and differences with other media (Section \ref{sec:properties}).
We show challenges in representing podcasts for downstream information access tasks in Section~\ref{sec:representation}, and highlight opportunities for further research in podcast representation.
We expand upon podcast consumption patterns, listening behaviors, and potential implicit feedback signals that can be used for estimating user satisfaction for training and evaluating podcast information access systems (Section~\ref{sec:consumption}). We further provide an in-depth perspective on research potential related to information access technologies in Section~\ref{sec:infoaccess}. They include podcast search, recommendation, and social discovery, in addition to podcast summarization, which is necessary to provide a preview of podcasts' spoken content to users. We also briefly touch on a aspects related to user experience in podcast information access, which is another under-explored facet for future research.

Together, this perspective paper will demonstrate that podcasts are significantly different from other spoken document corpora, and that we should not treat them as ``noisy text'' using pipelined approaches. Rather, podcasts should be handled using holistic approaches that take advantage of their multimodal  and hierarchical signals. This points to a future of podcast research integrating audio and text approaches, hierarchical and end-to-end models, and representing both listeners and creators.

\vspace{-0.2cm}
\section{Podcast Properties}
\label{sec:properties}

In this section we describe the unique properties of podcasts. 

\begin{figure}
    \centering
 \includegraphics[width=3in]{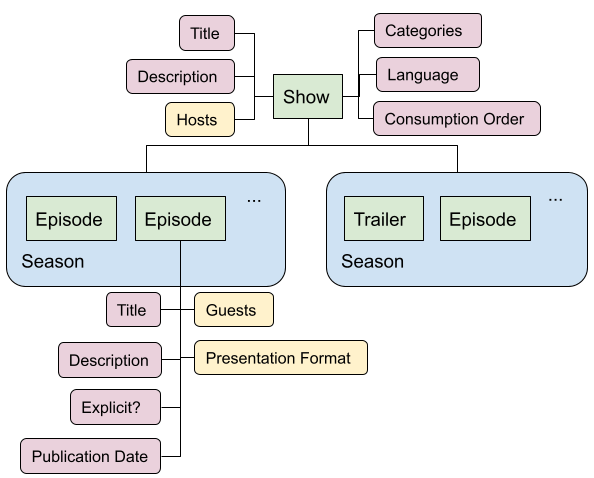}
    \caption{Structure of the metadata associated with a podcast. Metadata in pink is provided by the podcast creator in the RSS feed, while the yellow boxes are examples of properties that are not explicitly specified.}
    \label{fig:properties}
    \vspace{-0.4cm}
\end{figure}

\paragraph{\textbf{Structure and metadata}}
Podcasts are distributed as audio streams or files,
traditionally through RSS feeds.
The RSS standard for podcasts contains multiple metadata fields\cite{podcastersguiderss}.
Figure \ref{fig:properties} illustrates the hierarchical structure of a typical podcast, and some of the metadata that is associated with shows and episodes. 
A podcast {\em show} has a title, description, language, consumption order (episodic or sequential), and a list of categories (e.g., Society \& Culture, Sports, and Comedy) selected by the creator from a predefined taxonomy; the show is represented by the RSS feed.
A show typically comprises multiple {\em episodes}, which are the distinct audio files streamed or downloaded by a listener. Each episode has its own title, description, artwork, and other information.
Episodes may be organized into {\em seasons}, though this is generally an informal designation with no associated metadata. 


As noted in previous studies~\cite{sharpe2020review}, the metadata found in the RSS feeds is often noisy and inadequate.
Episode descriptions are of varying quality and scope. 
Category labels cannot be considered completely reliable: categories are ill-defined, and podcast creators are incentivized to list their shows under multiple categories to maximize exposure.
There is a research opportunity to identify and label categories which are meaningful to podcast listeners and automatically categorize podcasts into those categories.

\begin{figure*}
\centering
\vspace{-1.5cm}
\subfloat[Episode duration (up to 90 minutes).]{
    \includegraphics[width=0.65\columnwidth]{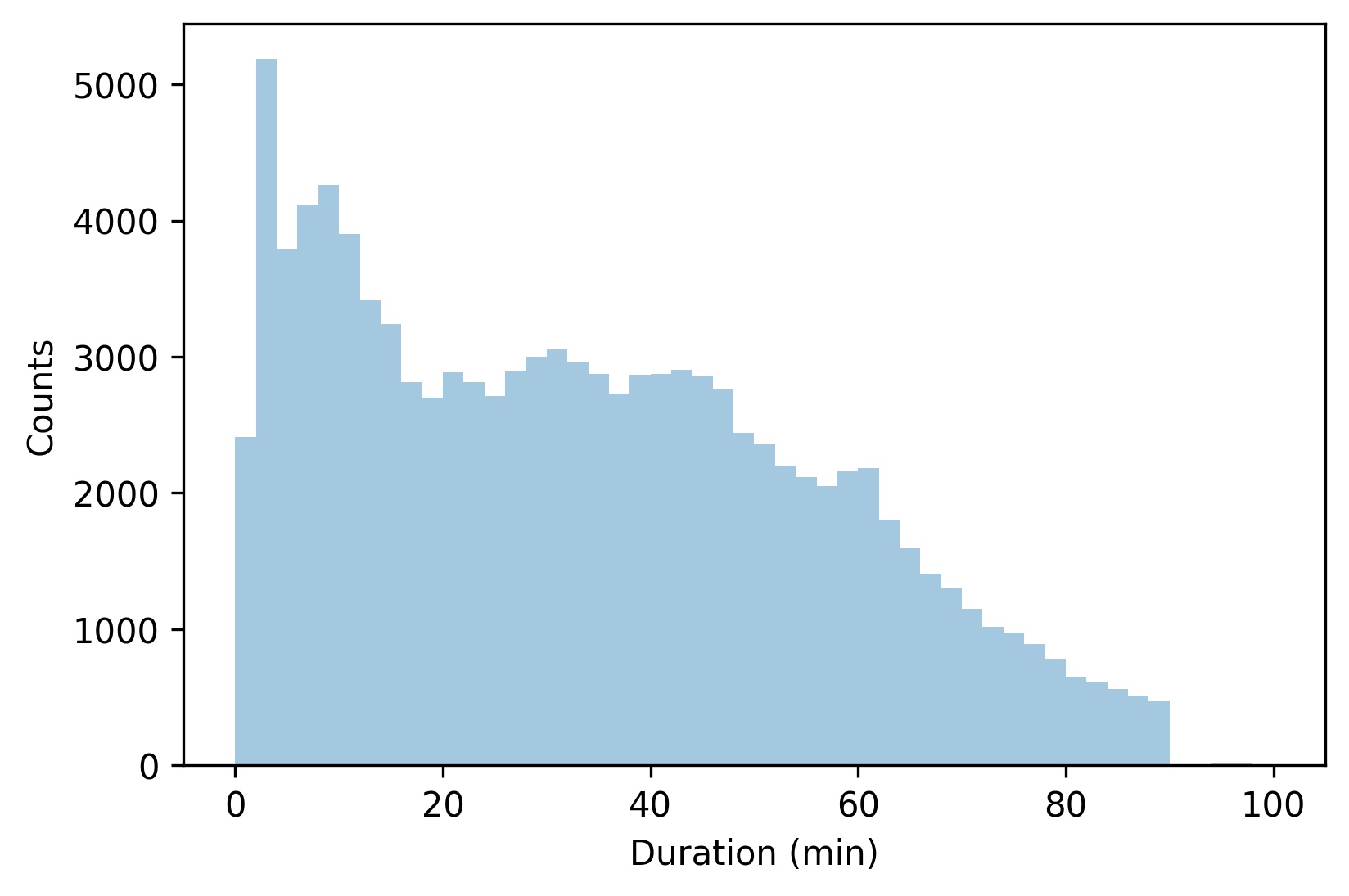}
    \label{fig:stats:duration}
    }
\subfloat[Number of speakers ]{
    \includegraphics[width=0.65\columnwidth]{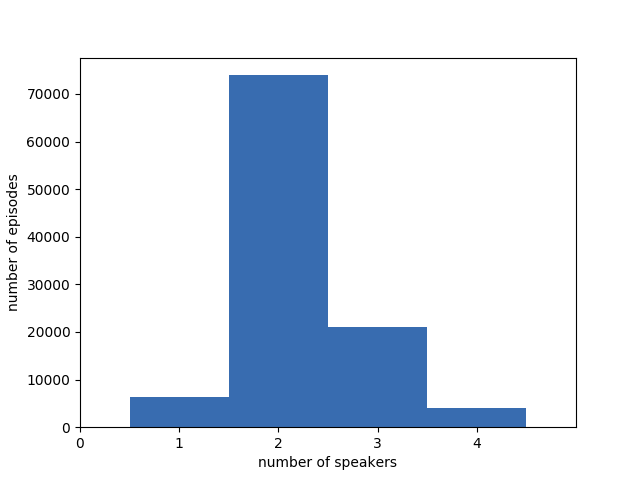}
    \label{fig:stats:numberofspeakers}
    }
\subfloat[Primary speaker's share of time speaking]{
    \includegraphics[width=0.65\columnwidth]{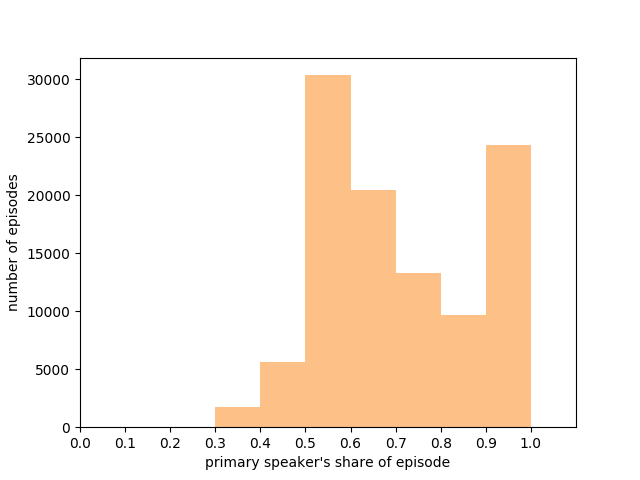}
    \label{fig:stats:speakershare}
    }
    \vspace{-.3cm}
    \caption{Distributions of episodes by duration, number of speakers, and share of primary speaker in the 100k English-language Podcast Dataset, from Clifton et al. \cite{Clifton:2020:COLING}. Speaker diarization is computed automatically, and while it may be noisy, the aggregate distributions demonstrate the different conversational styles in podcasts. 
    }
    \label{fig:stats}  
    \vspace{-.4cm}
\end{figure*}



\paragraph{\textbf{Format and content}}
Podcasting is similar in form and content to talk radio, in that it is typically a spoken-word medium.
However, the relative ease and low cost of recording and publishing means there is a great deal of variability in the specifics.

Podcast episodes have a wide range of lengths, from just a few minutes to hours.
On average, each podcast episode is between half an hour and an hour in length.
\figurename~\ref{fig:stats:duration} plots podcast duration in a research dataset of 100,000 podcast episodes released by Spotify ~\cite{Clifton:2020:COLING}. The span of episode lengths reflects a variety of use cases and situations for listening events.

Podcast episodes frequently contain mixed media: music, sound effects, and archival clips, as well as the recorded narration which frames the content. 
Since podcasts do not require visual attention, they enable a broader set of use cases than video material, similar to other audio media such as music or broadcast radio.

Speech
allows for a richer communicative channel than text, including dialectal, sociolectal, and individual variation, which all is normalized when language is written. 
Some podcast material is scripted, like some radio broadcasts and audiobooks.
Other podcasts consist of informal and unstructured dialogue, created with the audio as the primary channel. 
This poses a general challenge to most existing information access systems: they are built to use a written representation, and the representation of a podcast episode in writing removes much of what characterizes it. Some of the use cases we know podcasts are created for are likely to hinge on the identification of that specific variation, e.g., material produced in a local variety of a standard language. 

Presentation formats of podcasts vary widely, from monologs to multi-party conversations (see \figurename~\ref{fig:stats:numberofspeakers}); from lectures and narratives to interviews, sermons, debates, and chatty conversations; from newly recorded material to historical clips; from dispassionate discourse to argumentation, jokes, or rage.  

Moreover, podcast episodes are frequently anchored into a shared context between the creators and audience, and may require cultural familiarity to be fully understood. Some of these characteristics are likely to inform listening choice and selection, but information access systems today \emph{are not} equipped to classify material in non-topical categories without editorial oversight.

The range of style, format, and language variety, and the variation exhibited by new use cases, motivates research effort to model and understand usage in more detail and to identify how the variety of material can be mapped to the variety of usage through information access tools and technology. This means developing technologies for identifying and representing the variation and leveraging that variation into useful features for information access systems, both for classification and for direct presentation to users.

\vspace{-0.2cm}
\section{Podcast Representation}
\label{sec:representation}

How we choose to represent the information contained in podcast shows and episodes plays a key role in achieving efficient and effective information access. 
In this section, we describe several representations  suitable for podcast information access and discuss their shortcomings, as well as opportunities for research.

\paragraph{\textbf{Metadata}}
Due to the diverse nature of podcast properties discussed in Section~\ref{sec:properties}, which include both structured data and free text, we can use semi-structured or fielded document representations for podcasts.
Free-text fields such as title and description can be treated as simple bags of words, admitting distributed vector representations.


Though optional metadata attributes and incomplete facets introduce a number of challenges for such systems, this paradigm has traditionally been quite successful in multimedia information access, such as music. However, because the essence of the podcast lies in the actual spoken audio,
a deeper understanding of the podcast and its content could improve access. 

There is also an opportunity to take the hierarchical nature of podcasts into account, for example, a model which can represent an episode, taking into account both the the parent show and sibling episodes, and the structure of the episode, could lead to better information access, and we recommend research in this direction.

\paragraph{\textbf{Transcription}}
The main media component of podcasts is the audio stream. 
To enable content-based search, browsing, and recommendation, using traditional text-based approaches, a full textual representation in the form of a transcript is valuable.
Human-generated transcripts are expensive and not produced by typical podcast creators.
Instead, automatic speech recognition (ASR) can be used to infer a textual representation from the audio stream \cite{Yu2014ASR}, which could then be added to the fielded document representation.

Transcribed spoken content is fundamentally different from written text 
due to the lack of 
sentential and paragraphic boundaries, 
as well as spoken disfluencies.
Thus, spoken content is often indexed or organized differently in comparison with written text~\cite{chelba2008retrieval}. 
Noise due to ASR errors may be significant: 
the word-error-rate in the 100k Spotify Podcast Dataset is reported to be 18\% \cite{Clifton:2020:COLING}.
Though spoken collections of news have previously been studied \cite{eskevich2012new}, the news domain is much more constrained and leads to better accuracy than can be expected on the wide variety of genres, levels of professionalism and languages found in podcasts.
Research into ASR on the podcast domain could lead to lower error rates. In addition, using information from the podcast meta-data could lead to improved ASR performance. Another valuable research direction is end-to-end modeling, where ASR is implicitly built into the task, with audio input feeding into a deep neural model, with an information access task such as retrieval as the output.

\paragraph{\textbf{Acoustic features}}
As mentioned  in Section~\ref{sec:properties}, podcasts may contain several kinds of non-verbal audio content. Therefore, speech transcription alone leads to information loss and thus sub-optimal information access. To address this issue, one could enrich podcast representations using acoustic features such as MFCCs~\cite{barthet2010speech}, PLPs~\cite{ogata2012podcastle}, and more recently using ALPRs~\cite{yang2019more} that are more robust and suitable for podcasts. Such representations are effective and could leverage unlabeled data, however they are not interpretable with respect to downstream applications. 
We may also wish to derive interpretable features from the audio. Yang et al ~\cite{yang2019more}  showed they could use ALPRs to predict seriousness and energy of podcasts,  as well as popularity. Acoustic features take advantage of a unique aspect of podcasts, and can be used as part of a multimodal approach to podcast information access. 

\paragraph{\textbf{Semantic podcast representation}}
For human-consumable podcast browsing, 
knowledge bootstrapping and aggregation can be useful.
Semantic web techniques have been applied to podcasts to induce an RDF-like structure to the metadata and audio content~\cite{celma2008zempod}. 
A structured representation of the podcast metadata using knowledge graphs could also be computationally effective as has been found in  related multimedia such as news recommendation using \textit{NewsGraph}~\cite{liu2019news}, spoken content retrieval using semantic structures~\cite{lee2015spoken} and large scale video classification~\cite{abu2016youtube}. 

The representation of podcasts via a heterogeneous graph could help with analysis of the most important (or well-connected) nodes, and thus their effect on downstream applications. A knowledge graph (KG) is a multi-relational, directed heterogeneous graph, composed of entities (nodes) and relations of different types (edges)~\citep{ji2020survey}. KGs are used to power question answering~\citep{dong2015question}, semantic search~\citep{szumlanski2010automatically}, and more recently recommender systems~\citep{wang2019explainable,palumbo2017entity2rec, wang2018ripplenet}. 
Because there are few explicit connections between nodes in a podcast domain, we may rely on information extraction techniques such as entity set expansion~\cite{paulheim2017knowledge}, relation extraction~\cite{zeng2014relation}, and link prediction~\cite{kazemi2018simple} to enrich the heterogeneous graph. Such a graph could be leveraged to provide insights into similarity between podcast shows and explainable connections via edges connecting their entities. 
One could further learn semantic relations between nodes using embedding based methods that perform random walk along the graph~\cite{bordes2013translating} to produce latent podcast representations.

While there are challenges in transcription, as described above,
the redundancy between transcripts and descriptions could  mitigate ASR errors and help us perform named entity recognition and disambiguation on the free-form text. 
Such entities often include people such as hosts and guests.
In addition, 
entities such as genres, topics of discussion, and conversational styles are relevant elements of a podcast. This variety of entities could be interrelated and their interplay could result in personalized experience to the listeners. 

We recommend further research into the effectiveness of existing knowledge graph work on podcast information access, as well as augmenting entity extraction in a multimodal way from text descriptions and audio or transcripts of audio.

\vspace{-0.2cm}
\section{Podcast Consumption and Feedback}\label{sec:consumption}
Compared to many multimedia items, such as music, consuming podcasts requires a significant time investment. Therefore, it is particularly important to understand how users consume podcasts, as this can help us to identify implicit feedback that can be used for training and evaluation of podcast information access systems.

\paragraph{\textbf{Podcast consumption patterns}}
Podcast consumption patterns are influenced by the diverse user needs they serve, as well as show characteristics such as release frequency and average episode length. Podcast consumption can be studied at an aggregated user level.
A national survey conducted by Edison Research in the United States~\cite{Edison2019Consumer} revealed that the top four reasons or goals for listening to podcasts include: learning new things (74\%), entertainment (71\%), staying up-to-date with latest topics (60\%), and getting relaxed (51\%). This variety in user goals has created an ecosystem of podcast shows that are structured to address the diverse needs of the user. Informational shows that are aimed at keeping their listeners up-to-date may be released on a daily basis, while entertainment or true crime shows are often structured in seasons and released weekly, similar to TV shows. 
Consumption patterns in the podcast domain are different from other mediums. A study on podcast consumption patterns by \citet{li2020podcasts} demonstrated that users listen to podcasts during weekday mornings,
whereas users listen to music during evenings, nights, or weekends. However, consumption patterns are personal and user level podcast consumption has been left relatively unexplored. Such analysis would lead to more accurate personalized information access systems.

\begin{figure}
    \centering
    \vspace{-1cm}
    \includegraphics[width=3in]{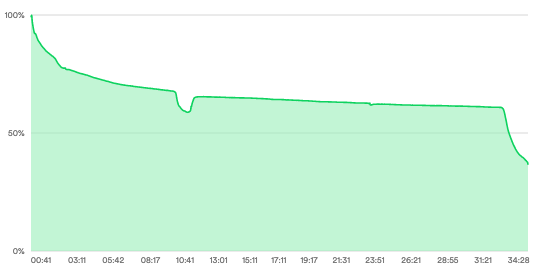}
    \vspace{-0.3cm}
    \caption{Illustrative curve showing the proportion of users that streamed each second of a single episode on the Spotify platform. Episodes tend to show the steepest drop-offs in listenership at the beginnings and ends, with some dips in the middle that mainly correspond to ads.}
    \label{fig:retentioncurve}
    \vspace{-0.4cm}
\end{figure}

\paragraph{\textbf{Listening curves}}
A feature of streaming media such as audio and video is that it is possible to measure listener attention at different points in time over the duration of the stream. While tracking attention is also possible with text documents on web browser,s
knowing the exact start and end points of a listener's stream (within the client instrumentation capabilities) is a more precise signal of attention. 
Figure \ref{fig:retentioncurve} is one illustrative example of a curve from listening data on Spotify that shows the proportion of listeners at each time point over the duration of a single podcast episode. In such curves, dips tend to correspond to ads or other extraneous material \cite{podcastaddetect}
and there are commonly sharp drops at the beginnings and the ends of episodes. These curves in general show some distinctive characteristics depending on the nature of the podcast; for example, well-known podcasts tend to have a sharper drop at the beginning than lesser known podcasts, since they attract a diverse group of listeners who may be curious about the podcast but find that they are not interested after a few seconds. Listening curves are useful for detecting extraneous content\cite{reddy-etal-2021-detecting},  assessing ad monetization,  improving summarization, and  devising user engagement metrics on podcast-access platforms (such as, for example, deriving thresholds on the amount of listening that counts as user satisfaction).

\begin{figure}[t]
    \centering
    \vspace{-1cm}
    \includegraphics[width=2.5in]{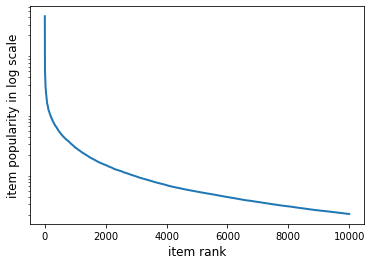} 
    \vspace{-0.3cm}
    \caption{The long-tail characteristic of the popularity distribution of the top 10000 podcast shows on Spotify. Popularity is measured as the absolute number of streams.}
    \label{fig:podpop}
    \vspace{-0.4cm}
\end{figure}

\paragraph{\textbf{Popularity bias}} 
Similar to other domains such as music~\cite{kowaldmusicpopularity} and movies~\cite{abdomoviepopularity}, a small number of podcast shows dominate the probability distribution as shown in \figurename~\ref{fig:podpop}. Therefore, careful treatment of the items in the long-tail is necessary not only to ensure user satisfaction \cite{steck2011} but also to guarantee diversity and fairness \cite{mehrotra2018fairness} in podcast information access systems.

\paragraph{\textbf{Podcast consumption as implicit feedback}}
On most platforms, users can subscribe to their favorite podcast shows and be notified when new episodes are released. Users may also  ``drop-in'' to  other shows and listen to specific episodes from a show due to their topic, guest or other reasons, without subscribing to the show. As in many other domains, eliciting explicit feedback from podcast listeners in scale is impractical. Therefore, inferring user satisfaction with a show or episode relies on implicit signals. In the podcast domain, {\it subscribing} provides a reliable form of feedback that shows the user's interest in a show. However, it fails to create a holistic picture of a user satisfaction with the interacted item. This means that {\it play}, {\it pause}, and {\it stop} are also important signals, specifically on the episode level. Such signals can then be interpreted in a variety of ways to estimate user satisfaction. For instance, total listening duration, number of pauses, and listening abandonment, can be used as implicit feedback signals. Due to the high variance of podcast episode lengths, some of these signals may need to be normalized based on length of the podcast episode. Identifying and characterizing each of these signals are important research questions that need further investigation.

Moreover, having multiple implicit feedback signals provides both opportunities and challenges. At times they appear to be contradictory -- users may subscribe to certain shows but  not listen to them. Aggregating various signals  can be challenging in inferring user satisfaction with a show or episode, and may be user-specific. 





\section{Podcast Information Access}
\label{sec:infoaccess}
Information access tools, such as search engines and recommender systems, are an essential part of finding and discovering podcasts. In this section, we highlight unique characteristics of information access in the podcast domain. We first review challenges in developing search engines and recommender systems for podcasts. We further discuss our perspective on social podcast discovery, for example through social media. We then study podcast summarization as an essential part of generating previews and trailers for podcast information access tools. We finish with covering user experiences with podcast information access systems.

\subsection{Podcast Search}
\label{sec:search}
Podcast search as shares qualities with several other search settings, while also having its own unique characteristics and challenges.  In particular, podcast search is related to:
\begin{enumerate}[leftmargin=*]
    \item \emph{semi-structured document retrieval}: as described in Section~\ref{sec:representation} podcasts can be represented as semi-structured documents and retrieval models, like BM25F~\cite{Robertson:2004}, Field Relevance Models~\cite{Kim:2012}, and NRMF~\cite{Zamani:2018:NRMF}, can be adopted for podcast search tasks. A number of evaluation campaigns, such as INEX XML retrieval initiative \cite{Fuhr:2006,Lalmas:2007}, have studied such models.
    \item \emph{spoken document retrieval}: podcasts can be  represented by their transcripts of their spoken content - in this way podcast search is related to spoken document retrieval \cite{garofolo2000trec},\cite{AkibaEtAl2013},\cite{PecinaEtAl2008}. 
    \item \emph{multimedia retrieval}: as pointed out in Section~\ref{sec:consumption}, entertainment is one of the second most important consumption goals in the podcast domain, which is in part similar to most multimedia items, such as music and movies. 
    \item \emph{blog search}: as argued by \citet{Besser:2008:UserGoalsPodcastSearch}, the underlying goals of podcast search may be similar to those for blog search, as podcast can be viewed as audio blogs. 
\end{enumerate}

Below we highlight novel aspects of podcast search and potential future research directions.

\paragraph{\textbf{Podcast search tasks}}
Perhaps the simplest search task in the podcast domain is {\em catalog search}, usually in the form of podcast show or episode title search.
Misspellings, forgotten identifiers, and other errors can make the task much more difficult for a search engine to complete.
For example, ``tip of the tongue'' search~\cite{Arguello:2021} is a case where a user has previously heard of or listened to an episode but cannot recall a reliable identifier.

As described in Section~\ref{sec:consumption}, users listen to podcasts for a variety of reasons, including entertainment, education, and information; these use cases can translate into search tasks
that require more than catalog match.
Podcast informational search tasks may be similar to traditional informational search, in which the user wants to find relevant information about a topic, but there are unique differences stemming from the variety of formats podcasts can take, and unique challenges due to the potential difficulty of finding relevant information in audio or noisy transcripts. 
Podcast segment retrieval was proposed and studied in the TREC Podcast Track~\cite{trec2020podcastnotebook} with informational and known-items queries. The goal was to retrieve a part of a podcast that is relevant to an information need. This is closely related to passage retrieval task with a heterogeneous collection.  Traditional text retrieval approaches can be applied to this task: for example, \citet{Clifton:2020:COLING} showed that term matching retrieval models, such as BM25~\cite{Robertson:2009:BM25} and query likelihood~\cite{Ponte:1998:QL}, can achieve an NDCG@5 greater than $0.25$ on a large-scale podcast search collection for a small set of queries with manual relevance annotations. 

Entertainment and education tasks benefit greatly from personalized search, to leverage knowledge of the types of content users find entertaining or educational respectively.
Indeed, personalized search seems to be almost a necessity for podcast search, due to varying user tastes for differing podcast formats, tolerance for low-quality audio, affinity for the hosts, not to mention contextual factors such as the window of time the user can devote to listening and what else they may be doing while they listen.


\paragraph{\textbf{The notion of relevance in podcast search}}
From an information science perspective, the concept of relevance lies at the convergence of understanding users, information needs, items of information, and interaction. Relevance --- the momentary quality of a text that makes it valuable enough to read --- is a function of task, text characteristics, user preferences and background, situation, tool, temporal constraints, and untold other factors and has in information retrieval evaluation been formalized to be a relation between a description of a user information need and documents or information items in a collection, generalizing over other contextual or individual factors and based on topical similarity \cite{mizzaro1997relevance,mizzaro1998many,belkin2011entertainme,belkin2012report,Saracevic:1975:Relevance}. 

The notion of relevance for catalog search is straightforward, and for straight informational search may be a relatively straightforward translation from traditional search tasks.
But because podcasts are often used simultaneously for entertainment, education, and information, the enjoyability and appeal facets of relevance are at the fore.
This argues for a personalized and contextual notion of relevance.
Personalization has been well-studied in web search~\cite{Bennett:2012,Matthijs:2011}, and some techniques such as leveraging past consumption history are likely to be useful for podcast search as well.
Contextual search has been less well-studied; work stemming from the TREC Contextual Suggestion track~\cite{TRECCS} may be the most relevant, though it focused on very specific geographical contexts.


In addition, the publication format of podcasts, as series of episodes typically consumed in sequence, and the prominence of hosts and certain popular guests, act as a filter on top of the topical search. \citet{Manos:2008:PodCred} argued that the quality and credibility of podcasts, which are sometimes considered during the relevance assessment process, can be characterized using four types of indicators pertaining to the podcast content, the podcast creator, the podcast context, or the technical execution of the podcast. These facts distinguish podcast search from most well-established search tasks including adhoc, web, personal, and enterprise search.



\paragraph{\textbf{Podcast collections}}
A podcast catalog consists of podcast metadata (RSS feeds) and audio and is constantly growing as new shows and episodes are added.
A podcast search collection is likely to be a snapshot of a catalog, constructed from the metadata and audio (likely via transcription). 
As pointed out in Section~\ref{sec:representation}, transcribing podcasts is not flawless and this may influence the retrieval performance. In addition, podcasts are often long, thus their transcriptions result in very long documents. For example, the average and maximum document length in the Spotify Podcast Dataset~\cite{Clifton:2020:COLING} are 5,728 and 43,504 words, respectively.
It is well-known that great variation in length creates challenges for standard IR models~\cite{buckley:sigir96}.

Moreover, podcast collections are heterogeneous. Some podcasts have one speaker, while some include conversation between multiple persons. 
In addition, podcasts may contain non-verbal information, such as music or background audio effect. 
How to incorporate heterogeneous, multimodal information into a search engine is potentially a rich line of research.

Additionally, podcast catalogs are dynamic.
They quickly evolve and grow --- and sometimes shrink when creators delete old episodes.
This calls for research on temporality and sequentiality in search.
In particular, understanding the relation between world events and new items in the catalog could be beneficial in podcast search tasks.

Since podcasts are user-generated, they are of varying quality and credibility. This fact should be considered in podcast search engine development. We cover the concept of trust and credibility in Section~\ref{sec:recsys}.

\paragraph{\textbf{Podcast search evaluation}}
Evaluating podcast search engines is challenging. 
Even after deciding on a definition of relevance, podcasts can be difficult to assess.
Whether assessment is done by listening or by reading transcripts, it is potentially expensive and time-consuming. 
Long podcasts may touch on a diverse set of topics and therefore require careful attention to determine relevance, particularly in errorful transcripts. The {\sc TREC} Podcast track \cite{trec2020podcastnotebook} addressed this by making the retrieval unit two-minute segments, so that each segment was relatively easy to assess by reading the transcript, and the relevance judgements could be re-used.

Podcast relevance is often personal and contextual, which makes reusable assessments far more difficult to collect.
Even when implicit feedback such as streams or subscriptions is available, it does not necessarily indicate relevance. Inferring the relationship between context, relevance, and observed implicit feedback in podcast search is challenging and requires further investigation.

With relevance assessments, traditional IR evaluation measures like precision, recall, and nDCG can be used for evaluating podcast search.
Here too there is opportunity for novel research on new metrics.
For example, metrics that account for the amount of time it takes to consume retrieved content could prove useful~\cite{smucker:sigir12}.

\paragraph{\textbf{Result presentation}}
Search engines often provide a short preview or summary of the retrieved items for quick assessment by users, so they can faster find the relevant items. For instance, snippets are used in web search for summarizing documents. 
In the context of podcasts, episode descriptions can be employed as a summary, however, as mentioned in Section~\ref{sec:properties}, this field is optional and of varying quality. 
Alternatively, search engines could display a transcription of a relevant snippet of the audio, or more generally, automatically generate a query-biased summary of the episode.
This is challenging; we discuss in more detail in Section~\ref{sec:summarization}. 

\subsection{Podcast Recommendation}
\label{sec:recsys}

Podcast recommender systems have been categorized as speech recommenders~\cite{10.1145/3407190}, which
falls short of capturing their multimodal nature discussed in Section~\ref{sec:properties}.
While generic recommender system algorithms, such as variants of collaborative filtering, are applicable to this domain, the specific nature of podcasts---in particular their representation as (noisy) text through automatic speech recognition---makes content-based and hybrid approaches appear particularly well-suited. In fact, the few published approaches to podcast recommendation typically leverage textual information attached to the audio~\cite{DBLP:conf/bigdataconf/XingPXKP16} or extracted through conversational interfaces~\cite{DBLP:conf/recsys/YangSTE18}.
A very recent approach is trajectory-based podcast recommendation, which models short-term listening behavior of users as trajectory in a podcast graph, and predicts the next shows a user is likely to access~\cite{DBLP:journals/corr/abs-2009-03859}.
More precisely, this sequential approach represents a given collection of podcasts as a graph where nodes are shows and edges connect them through shared topics. 
Enriched with podcast descriptions (e.g., from Wikipedia), a graph embedding approach is applied, which represents nodes in a semantic embedding space. Users are then modeled as temporal sequences of these node embeddings, and a recurrent neural network architecture (LSTM-based) is used to predict the most likely next show the target user may listen to.
While existing work has yielded interesting insights and first results, major open questions and challenges need to be addressed and investigated in depth. 

\paragraph{\textbf{Cross-domain recommendation}}
Since it is often music streaming platforms that extend their catalogs to include podcasts, a natural choice to address the cold-start problem (in this case, missing initial user-podcast interactions by existing users), is to adopt a cross-domain recommendation approach. In particular, using music preferences in a cross-domain fashion to address cold-start in podcast recommendation has been shown to be successful~\cite{DBLP:conf/sigir/NazariCPLCVC20}.
On the other hand, domains such as movies or books have not yet been investigated for cross-domain podcast recommendation. Neither have microblogs or other (textual) user-generated content shared on social media, which presumably hold rich information about topical preferences of users.

\paragraph{\textbf{Duration-aware recommendation}}
Compared to music, where individual recordings have a duration of typically a few minutes, the duration of episodes in podcasts spans a wide range, from a few minutes to several hours (see \figurename~\ref{fig:stats:duration}). 
Furthermore, while it typically takes a listener only a couple of seconds to decide whether or not they like a song, this time is much longer for podcasts, in fact may be closer to the time needed for a movie or TV show.
Poor podcast recommendations are therefore likely to more severely deteriorate user experience than poor recommendations in the music domain.
The implications these characteristics have on the ways users interact with items, and consequently the requirements of a podcast recommender system, are still open to investigation.

\paragraph{\textbf{Tailoring recommendations to situational context}}
Unlike music, which is frequently consumed in the background, podcasts are almost exclusively consumed in active listening modes, i.e., listeners pay close attention to the content.
In addition, podcasts are often consumed while commuting~\cite{digitalnewsreport:podcasts}.
As a result, temporal constraints are typically more pronounced for podcasts than for music. Because podcasts are much longer than songs (see Fig.~\ref{fig:stats:duration}) and users tend to prefer to listen to them sequentially, both situation and timing need consideration. A user on a 30-minute-commute may prefer podcasts of approximately this duration over significantly shorter or longer recommendations.
The influence of the listeners' situational context and temporal constraints while consuming podcasts on user behavior and preferences, as well as their implications on algorithm design and system evaluation has not been investigated yet - we recommend further research in this area. 
Nor is it known to what extent standard context-aware recommendation algorithms can be used or need adaptation to suit the domain of podcasts. 




\paragraph{\textbf{User characteristics and podcast preferences}} 
Extensive research has already been conducted to uncover relationships between a wide range of user characteristics and preferences for various kinds of media items; the former including user demographics \cite{Rentfrow2003DoReMi,DBLP:conf/recsys/FerwerdaTS17}, inclination to mainstream vs. niche items~\cite{DBLP:journals/corr/abs-1912-06933,DBLP:conf/ismir/VigliensoniF16}, and personality traits \cite{doi:10.1177/0305735610388897,10.1145/3340631.3394874}; the latter including movies~\cite{Golbeck:2013:PMP:2492517.2492572}, books~\cite{Rentfrow_etal:JP:2011}, and music~\cite{Ferwerda:2017:PTM:3079628.3079693}.
User models encoding such relationships can be integrated into recommendation algorithms and used to tailor recommendations to certain user groups, to mitigate cold start, or to diversify recommendations. 
Due to the recency of the podcast recommendation task, similar studies on how user characteristics shape preferences for podcasts do not exist yet. 
Also, we do not know yet whether inferred user models are capable of improving podcast recommendation performance similar to other domains.
While a first study found that a podcast recommendation approach performs better when adopting a pre-filtering strategy with respect to the user's age~\cite{DBLP:journals/corr/abs-2009-03859}, more in-depth investigations of user models, personalization strategies, and their impact on podcast recommendation performance are required to assess the potential of insights (and their formalization) gained in the studies mentioned.

\paragraph{\textbf{Trust and credibility}}
While many podcasts are consumed as a form of entertainment,  informational or educational podcasts are common, and may contain opinion or commentary on current affairs. Because of this, it is important 
to consider the trustworthiness of different sources, and the safety of the listener. 
The concept of trust and credibility~\cite{flanagin2000perceptions,ginsca2015credibility} has been studied in information retrieval, though much of that research has focused on the perceived trustworthiness of the information source~\cite{lynch2001documents}. 
More recent research has focused on online misinformation~\cite{qazvinian2011rumor} and identification of fake news~\cite{lazer2018science,zhou2020survey}.
The credibility of podcasts in particular has been studied by~\citet{Manos:2008:PodCred}, analyzing the factors which affect listener's perception of podcasts' credibility, including characteristics such as production quality and speaker style. 
As podcasts become a more popular source of information on current events,  concerns over misinformation and ``echo chambers'' become more important. 
For instance, to what extent do podcast recommender algorithms reduce or amplify misinformation?
Do podcast services have the ability to identify and act upon content that contains misinformation?
How can podcast recommender systems reduce the risk of creating ``echo chambers'' which guide listeners further down paths of potentially dangerous misinformation?

\paragraph{\textbf{Preference elicitation and evaluation}}
Podcasts are serial, periodic media since new episodes are released on a recurring basis. Therefore, users interact differently with podcasts than with other media types such as music or movies.
As a result, implicit preference feedback differs.
In particular, users can subscribe to shows and follow podcast creators, indicating they want to know when the next episode is released or their favorite creator creates a new show or is featured in an episode, respectively. 
Whether a user subscribes to a podcast feed has been found to depend on a variety of factors, some of the most important being the length of the description, keyword count, whether the feed has a logo, episode duration, author count, and feed period~\cite{DBLP:journals/jasis/TsagkiasLR10}.
Feedback is also available at the level of individual episodes, for example starting, pausing, skipping after a certain time.
All these different kinds of feedback on (inter-related) items make interpretation of preference signals a challenge, and may even yield contradicting preference indications, for instance if someone likes a particular episode of a show or a creator appearing in an episode, but not the show in general.

These particularities of implicit feedback signals 
call for revisiting evaluation metrics and discussing whether we should adapt existing or devise new metrics that consider the different interaction types (e.g., subscribing, following, watching) and levels of feedback (e.g., creator, show, season, episode).
Another aspect relevant to evaluating podcast recommendation stems from the fact that listening to podcasts is more time consuming than, for instance, reading a tweet or listening to a song (see above). Therefore, the negative effect of a poor recommendation on user experience and retention is higher than for many other recommendation domains. Evaluation approaches could consider this, e.g., by giving higher priority to maximize precision and minimize false positive rate.

\subsection{Social Podcast Discovery}
\label{sec:social}
Like news readers \cite{Pew2019Facebook}, podcast listeners find content to consume outside of the context of algorithmic search and recommendation.
While searching the internet is the most common way to find podcasts (77\% of podcasts listeners do this occasionally), 67\% of listeners  find podcasts through  through social media posts, and 66\% find out about them through recommendations from friends and family~\cite{Edison2019Consumer}.
Other approaches to podcast discovery include non-podcast platform advertisements (web search/TV/radio ads), as well as cross-podcast recommendation (62\%) and advertising (54\%). Here we focus on social media discovery, due to its research potential.

We categorize social media discovery into two types. The first is when a listener's interest in a podcast is triggered by another user in a social media sharing information about a podcast. Since podcasts involve significant investments in time and energy, users need transparency about why a podcast is  shared, to engage with the podcast. Platforms should allow users to be more specific about what they find attractive about a podcast, by enabling them to share a particular quote from a podcast, its summary, a snapshot of a conversation, etc. to be able to interest more users to engage with the podcast. In Section \ref{sec:summarization} we elaborate on research directions for summarization and trailer generation.

The second category is a system making a recommendation to a user based on their trust network's preferences, also known as social recommendation~\cite{kautz:1997:commun}. 
\citet{king:2010:www} referred to social recommendation as any recommendation with online social relations as an additional input. Social relations can be interpreted as trusted relations, friends, or followers \cite{tang:2013:soc}. According to this definition, social recommendation systems assume that user preferences are correlated if they establish social relations. This is in contrast to traditional recommender systems which assume users are independent and identically distributed (i.i.d.\ assumption) \cite{ma:2011:acmtrans}. This assumption also makes sense intuitively. In the physical world, users often seek recommendations from friends, family, and generally their trust network. \citet{weng:2010:wsdm} shows this is also the case in social media. The authors have shown that users with {\it follow} relationship are more likely to share similar interests compare to two random users. 

The heterogeneous nature of social media means that different types of social relationships may have different impacts on social recommendation systems. For example user $u$ might trust user $u'$ about computer science, but not political topics. Using the trust relation that benefits social recommendation plays a particularly vital role when it comes to podcast media, considering their multi-faceted nature. Podcasts could have different providers, in different styles (e.g., interviews, storytelling,etc), with different hosts/guests. Even the podcast topic might change during a podcast show. All these new angles make podcasts intrinsically complex mediums. As a result, there are many problems that need to be revisited in  modern sociology such as online trust \cite{tang:2012:wsdm,Massa:2013}, community detection \cite{tang:2010,Leskovec:2010:www}, and heterogeneous networks \cite{sun:2012:book} in podcast recommendation.

\subsection{Podcast Summarization}
\label{sec:summarization}

Given that podcasts can be 30 minutes or longer per episode and the opening content does not always describe what is to come, listeners have to make an informed decision on whether the podcast is worth their time. 
Surveys reveal that listeners pay particular attention to the text description of a podcast in deciding whether to listen  \cite{mclean_2020}. The production of audio trailers is also a way for listeners to get a preview of the podcast. 
The composition of informative, accurate, and catchy descriptions and trailers is a time-consuming task; many 
podcasts have brief, uninformative descriptions and most have no audio trailers. We see an opportunity for automatic summarization of podcasts to serve information access needs in this domain, analogous to the role of summarization of long text documents such as news stories and research literature today. 

Podcast summarization is a new research area that recently got its start through the TREC 2020 Podcast Track \cite{trec2020podcastnotebook}. There are known challenges presented by speech for summarization that apply to podcasts: 
the ambiguity of utterance boundaries, natural conversational disfluencies, lack of explicit formatting, and errors introduced by automatic speech recognition \cite{mckeown2005text}. It also has unique challenges that make it distinct from traditional text or speech summarization, outlined below. 



\paragraph{\textbf{Abstractive versus extractive summarization}}
Because of the errors and disfluencies in transcribed speech, an abstractive model is best suited for a written text summary (although such a model may contain extractive components). Indeed, high-performing systems in the TREC 2020 podcast summarization task all produced abstractive text summaries \cite{trec2020podcastnotebook}.
On the other hand, a model to generate audio trailers could be primarily extractive; ideally, such a model should pay attention not only to the transcribed text content but also audio cues to indicate inclusion in a trailer. Trailer generation may also benefit from abstractive models that produce text to be read by voice actors or text-to-speech systems.  

\paragraph{\textbf{Multimodality}}
The generation of text summaries from podcast audio files presents a challenge, which can be addressed either through pipelined approaches that first use automatic transcription to convert the audio to transcribed text, and then transcription to summary, or through approaches that integrate the audio signal into the summarization model. 
While spoken language summarization has previously been studied  \cite{DBLP:journals/corr/ChenLCW16}, podcasts present additional challenges in the spoken domain. 
Above the previously seen challenges of noisy speech transcription and sentence segmentation, podcast often contain rapid, casual speech that compound these problems.
Further, podcasts are often voiced by many speakers without well-defined turn taking, and make use of nonlinguistic audio cues that are ignored by transcription systems.
Future approaches towards podcast summarization should leverage the audio signal directly to avoid spuriously impoverishing the model and propagating transcription error.

\paragraph{\textbf{Genre and use case diversity}}
A summarization system for podcasts must be able to handle a full array of styles and formats described in Section~\ref{sec:properties} and be robust to differences in speaking style, clarity, and structure of the content. 
In addition, podcast summarization calls for robustness to the style and use case of the summary. 
Traditional document summaries are designed to capture the key information of the documents such that a reader does not need to delve into the documents themselves if they are simply seeking to learn a few high-level takeaways. 
Podcasts, on the other hand, may be focused on non-informational subject matter that is more difficult to quantify informatively.
Accordingly, the role of summaries for podcasts is less clear cut, and may be informational or promotional. 
Most podcasts are designed to be consumed as audio experiences in their full forms, with factual information being no more important than the opinions, arguments, chit-chat, music, and creative structuring of the podcast. 


Many podcasts have promotional summaries, with catchy lines or hooks to entice listeners without giving away too much of the content.
On the other hand, some use cases for informational summarization may call for the generation of an outline-style summary.
Users may wish to consume particular segments within a podcast such as a question and answer, debate, or interview portion or other structural feature.
This differs from the search task in that it is not query based but rather assumes the discovery of user- and query-independent cohesive segments within an episode than may be defined by structural or format properties rather than by topic.
This type of summary should ideally automatically detect the chapter boundaries corresponding to the segment changes within a podcast.

Systems to produce automatic summaries must make  deliberate choices about the role of a summary.
While supervised models trained on a representative set of examples may implicitly learn to generate genre-appropriate summaries, they are not guaranteed to do so. These are considerations that should be accounted for when designing a general purpose podcast summarization system.

\paragraph{\textbf{Contextualization}}
For podcasts that are serialized, with each episode building upon the previous ones, creators may choose to include a recap of the previous episodes to establish context. Summarization systems could be designed for this use case, either to aid creators in composing such recaps or exposing automatically generated summaries of previous episodes to listeners. Such systems should take into account not only the episode being recapped, but the context of the episode which the summary accompanies. 

Another type of contextualization is producing personalized summaries which are tailored to a listener's  preferences, history, or a specific query. Personalized summarization has been explored in other media \cite{hannon2011personalized,li2019towards} but is an unexplored avenue for podcasts.


\paragraph{\textbf{Evaluation}}
Evaluation presents a significant challenge for podcast summarization. 
Already, there can be any number of ways to summarize a text, and generating even a single high-quality reference summary is expensive. 
These issues are exacerbated in podcast summarization, because of the wide variety of podcast genres and summary use cases, making it more difficult to define and quantify quality.
This is compounded by the exaggerated compression ratio of podcasts to summaries in comparison to the traditional summarization task, since traditional task documents, e.g. news articles, are typically much closer in length to their reference summaries than podcasts. 
Initial results  \cite{Clifton:2020:COLING,trec2020podcastnotebook} indicate that ROUGE metrics using podcast episode descriptions as the reference correlate weakly with expert human judgements, but future work should examine this more thoroughly, because the highly subjective and generative nature of producing summaries for podcasts is likely to exacerbate know issues with lexical matching metrics such as ROUGE \cite{peyrard-2019-studying}.
Additionally, research into reference-free task-based evaluation will be valuable for podcast summarization, particularly in light of the varied summarization use cases for the podcast modality.




\subsection{User Experience} 
\label{sec:userexp}


Improving the user experience of podcast information access possess unique challenges and opportunities. In this section, we highlight four distinct aspects that future research should invest in: information access interfaces, fine-grained information access, intentional information access, and multi-sided markets.

\paragraph{\textbf{Information access interfaces}}
People access podcasts  mostly through visual interfaces on mobile and desktop. However, the increased popularity of voice interfaces on smart devices  provides new channels for access, especially under hands-free scenarios like driving, cooking, etc. As discussed in Section~\ref{sec:infoaccess}, podcast search and recommendation additionally rely on rich metadata which is presented to the user to help them select from a range of options. When giving the listener the results of search or recommendation through a narrow voice channel, it can be  inefficient to deliver and navigate through the same meta-data~\cite{DBLP:conf/recsys/YangSTE18}. Future research can develop and leverage audio summarization (as also explored in Section~\ref{sec:summarization}) and text-to-speech methods to synthesize verbose information, and may also explore emerging hybrid interfaces (e.g., car displays) and multi-device settings. 

\paragraph{\textbf{Fine-grained information access}}
People find and discover podcasts at multiple granularity levels---a show, an episode, or a snippet within an episode. Finer-grained discovery presents new research challenges, including modeling nuanced user interactions within an episode and understanding the surrounding context in which the discovery happens: while some snippets can be consumed standalone, others may require additional background knowledge (e.g., news) or consumption order (e.g., true crime). Experimentation on retrieval of two-minute segments takes us a step in this direction \cite{trec2020podcastnotebook}, but future research should look at variable length segments as has been done on news documents \cite{eskevich2012new}. The user experience of snippets is an important area for future research as well.

\paragraph{\textbf{Intentional information access}}
Podcast consumption has been based on RSS subscription---people subscribe to the shows they plan to listen to and then consume the released episodes regularly. This characteristic makes podcast information access intentional. A field study~\cite{yang2019intention} has shown that overlooking listener intentions in podcast recommendations discourages listeners from achieving their aspirations and results in lower satisfaction. Future work should investigate mechanisms to elicit or infer  intentions in a light-weight way and incorporate such information into the discovery service.

\paragraph{\textbf{A multi-sided ecosystem}}
\label{multi-sided}
There are multiple sides in information access markets \cite{mehrotra2018fairness}, as both listeners and creators have a stake in how items are ranked and recommended. Podcasts also have advertisers as stake-holders, since podcasts frequently contain advertisements, both read by the podcast host, as well as those produced by the advertiser and inserted into the audio stream.
Podcast information access technology,
is mostly used by listeners.
However, such technology can also provide useful suggestions, in the form of metrics and consumption patterns, to  creators, to help them improve their content or reach a broader audience, as well as feedback for advertisers. This introduces some interesting challenges, such as satisfaction and fairness on all sides in information access. In the podcast context, research is needed to understand the trade-offs for multiple parties.
Future research can also explore designs that allow creators to customize their content for different interface (e.g., providing summaries for voice interface) and annotate their episodes for snippet-level podcast preview and listener matching.


\section{Summary}
\label{sec:future}

Our recommendations on
challenges and open directions in podcast research are focused around the need to reexamine typical textual
methods for the podcast domain and its novel use cases, and to leverage multiple channels of information. This includes intra- and inter-podcast structural organization and metadata, user information and listening behavior, and both linguistic and paralinguistic audio features of podcast content. 
In particular, for podcast \textbf{\textit{representation}}, future work should focus on developing unified representations over these multiple channels, for both generic and task-specific learning. Podcast \textbf{\textit{consumption and feedback}} has unique and challenging characteristics that should make it an area of focus. For podcast \textbf{\textit{search}}, research should account for the wide variety of podcast consumption goals to develop appropriate notions of relevance and personalization. Podcast \textbf{\textit{recommendation}} calls for cross-domain approaches, as well as duration-aware methods that leverage user context. The listening investment required of users by podcasts calls for the investigation into the process of \textbf{\textit{social podcast discovery}}, via networks and their specific technologies. For podcast \textbf{\textit{summarization}}, future work should be robust to the wide variety of genres and use cases, both in modeling and evaluation, and should recruit multimodal information. Finally, we advocate for research into \textbf{\textit{user experience for podcast information access}}, to gain understanding into user and creator needs in different interfaces and interaction types and their impact on experience. 



\section{Acknowledgements}
This work was supported in part by the Center for Intelligent Information Retrieval. Any opinions, findings and conclusions or recommendations expressed in this material are those of the authors and do not necessarily reflect those of the sponsor.

\bibliographystyle{ACM-Reference-Format}
\bibliography{main} 

\end{document}